\documentclass[preprint,12pt]{elsarticle}

\usepackage{listings}
\usepackage{color}

\definecolor{dkgreen}{rgb}{0,0.6,0}
\definecolor{gray}{rgb}{0.5,0.5,0.5}
\definecolor{mauve}{rgb}{0.58,0,0.82}

\usepackage{tikz}
\usetikzlibrary{backgrounds}
\usepackage{graphicx}
\usepackage{bm}
\usepackage{amsmath}
\usepackage{setspace}
\usepackage{hyperref}
\usepackage{subfigure}
\usepackage{multirow}
\hypersetup{
    colorlinks=true,       
}
\usepackage{amssymb}

\usepackage{lineno}
\usepackage{diagbox}

\newcommand{\calP}{\mbox{$P$}}
\newcommand{\calE}{\mbox{${\cal E}$}}
\newcommand{\calD}{\mbox{$D$}}
\newcommand{\calG}{\mbox{${G}$}}
\newcommand{\calZ}{\mbox{${\cal Z}$}}

\biboptions{sort&compress}

\journal{Journal Name}

\begin{document}



\title{Deep Learning of Turbulent Scalar Mixing} 



\author{Maziar Raissi$^{1}$, Hessam Babaee$^{2}$, and Peyman Givi$^{2}$}
\address{$^{1}$Division of Applied Mathematics, Brown University,\\ Providence, RI, 02912, USA\\
$^{2}$Department of Mechanical Engineering and Materials Science, University of Pittsburgh,\\ Pittsburgh, PA, 15260, USA}

\begin{abstract}
Based on recent developments in \emph{physics-informed deep learning} and \emph{deep hidden physics models}, we put forth a framework for discovering turbulence models from scattered and potentially noisy spatio-temporal measurements of the probability density function (PDF). The models are for the conditional expected diffusion and the conditional expected dissipation of a Fickian scalar described by its transported single-point PDF equation. The discovered model are appraised against exact solution derived by the amplitude mapping closure (AMC)/ Johnsohn-Edgeworth translation (JET) model of binary scalar mixing in homogeneous turbulence.
\end{abstract}

\begin{keyword}
PDF models \sep turbulent mixing \sep turbulent closure problem \sep system identification \sep data-driven scientific discovery \sep machine learning \sep predictive modeling \sep nonlinear dynamics \sep big data \sep high performance computing
\end{keyword}


\maketitle

\section{Introduction}

The problem of turbulent scalar mixing has been the subject of widespread investigation for several decades now \cite{Obrien60,Brodkey75,Pope00,Haworth10}. The problem is explicitly exhibited in the transported probability density function (PDF) description of turbulence in Reynolds-averaged Navier-Stokes (RANS) simulations. With the single-point PDF descriptor, the effects of mixing of a Fickian scalar appear in unclosed forms of the conditional expected dissipation and/or the conditional expected diffusion terms \cite{Pope00}. A similar closure problem is encountered in large eddy simulation (LES) via the probabilistic filtered density function (FDF) \cite{NNGLP17}. Development of closures for these terms has been, and continues to be, an area of active research; see \emph{e.g.} Refs.\ \cite{FFR04,Haworth10,AJSG11,Pope13} for reviews. The overarching goal of turbulence modeling is to find accurate closures for the unclosed terms that appear in PDF/FDF transport equations. As a common practice in turbulence modeling, the unclosed terms are formulated versus closed terms. The form of this closure is based on physical inspection of the problem at hand and it is inherently error prone. This is the major source of modeling uncertainty in turbulence closure.

In this paper, we introduce a new paradigm for turbulent scalar mixing closure, in which the unclosed terms are learned from high-fidelity observations.  Such observations may come from  direct numerical simulation (DNS) {\it e.g.} \cite{GZ96a,JMMG96ali,CD94} or space-time resolved experimental measurements, {\it e.g.} \cite{Pavlos_PIV,Gharib_DPIV}. Obviously, in  DNS, the unclosed term can be extracted  directly from the  simulated results. However, for most realistic applications, performing DNS is cost prohibitive.  On the other hand, finding closure from experimental data involves taking  derivatives in space-time and decomposition space from the experimental data (in some cases high-order derivatives), which is nontrivial and, even if possible,  introduces new uncertainty on the closure depending on the space-time resolution of  the measurements.   Our ultimate goal is to develop a closure discovery framework that learns the closure from sparse high-fidelity data, such as experimental measurements. The proposed framework replaces the \emph{guessing} work often involved in such  model development with a data-driven approach that uncovers the closure from data in a systematic fashion. Our approach draws inspiration from the early and contemporary contributions in deep learning for partial differential equations \cite{psichogios1992hybrid, lagaris1998artificial, sirignano2018dgm, weinan2017deep, long2017pde, baymani2010artificial, chiaramonte2018solving} and data-driven modeling strategies \cite{rudy2017data, rudy2018data, pan2018data}, and in particular relies on recent developments in \emph{physics-informed deep learning} \cite{RAISSI2018PINNs} and \emph{deep hidden physics models} \cite{JMLR:v19:18-046}.

As a demonstration example, we consider the binary scalar mixing which has been very useful for PDF closure developments \cite{Dopazo73,JKK79,Pope82,KG87,GMc88b,NP91,Girimaji92b,Girimaji92c,JG95,SP98,Pope13a}. The problem is typically considered in the setting of a spatially homogeneous flow in which the temporal transport of the scalar-PDF is considered. In this setting, development of a closure which can accurately predict the evolution of the PDF is of primary objective. The relative simplicity of the problem makes it well suited for both DNS and laboratory experiments. The literature is rich with a wealth of data obtained by these means; see {\it e.g.} Refs.\ \cite{GZ96a,JMMG96ali,TC81a,EP88,MCG89,CD93,CD94,SG91a,TV92,JW91,JW92}. We will demonstrate that our proposed framework rediscovers the conditional expected dissipation and diffusion.


\section{Binary Scalar Mixing}

We consider the mixing of a Fickian passive scalar $\psi$ = $\psi(t, \bm{x})$ ($t$ denotes time and $\bm{x}$ is the position vector), with diffusion coefficient $\Gamma$ from an initially symmetric binary state within the bounds $-1 \le \psi \le +1$.  Therefore, the single-point PDF of $\psi$ at the initial time is $\calP (0, \psi) = \frac{1}{2} [\delta(\psi-1) + \delta (\psi+1)],$ where $\psi$ denotes the composition sample space for $\psi (t, \bm{x})$. Thus $<\psi>(t=0) = 0$, $\sigma^2 (0) = 1$, where  $<\ >$ indicates the probability mean (average), and $\sigma^2$ denotes the variance.  In homogeneous turbulence, the PDF transport is governed by
\begin{equation}
\frac{\partial \calP}{\partial t} + \frac{\partial^2 ( \calE \calP)}{\partial \psi^2} =0,\ \ \ -1 \le \psi \le +1,
\label{eq:dissipation}
\end{equation}
where $\calE$ represents the expected value of the scalar dissipation $ \xi \ = \Gamma \nabla \psi \cdot \nabla \psi$, conditioned on the value of the scalar
\begin{equation}
\calE=\calE (t, \psi) = <\xi \vert \psi(t, \bm{x})=\psi >,
\label{4}
\end{equation}
where the vertical bar denotes the conditional value. Equation (\ref{eq:dissipation}) is also expressed by
\begin{equation}
\frac{\partial \calP}{\partial t} + \frac{\partial ( \calD \calP)}{\partial \psi} =0,
\label{eq:diffusion}
\end{equation}
where $\calD$ denotes the conditional expected diffusion
\begin{equation}
\calD=\calD (t, \psi) = <\Gamma \nabla^2 \psi \vert \psi(t,\bm{x})=\psi >.
\label{6}
\end{equation}
The closure problem in the PDF transport is associated with the unknown conditional expected dissipation, $\calE$, and/or the conditional expected diffusion, $\calD$.  At the single-point level none of these conditional averages are known; neither are their unconditional (total) mean values
\begin{equation}
\label{Total_Dissipation}
\epsilon(t)=\int_{-1}^{+1} \calP(t,\psi)  \calE(t,\psi) d \psi = - \int_{-1}^{+1} \psi \calP(t,\psi)  \calD(t,\psi) d \psi.
\end{equation}
%

\section{Deep Learning Solution} 

Given data $\{t^n, \psi^n, P^n\}_{n=1}^N$ on the PDF $P(t,\psi)$, we are interested in inferring the unknown dissipation $\calE(t,\psi)$ and diffusion $D(t,\psi)$ by leveraging Eqs.\ \eqref{eq:dissipation} and \eqref{eq:diffusion}, respectively, and consequently solving the closure problem. The data may be obtained from DNS or experimental measurements. 

\subsection{Conditional Expected Diffusion}

Inspired by recent developments in \emph{physics informed deep learning} \cite{RAISSI2018PINNs} and \emph{deep hidden physics models} \cite{JMLR:v19:18-046}, we propose to approximate the function
\begin{equation*}
    (t, \psi) \longmapsto (P, D)
\end{equation*}
by a deep neural network taking as inputs $t$ and $\psi$ while outputting $P$ and $D$. This choice is motivated by modern techniques for solving forward and inverse problems involving partial differential equations, where the unknown solution is approximated either by a Gaussian process \citep{raissi2018numerical,raissi2018hidden,raissi2017inferring,raissi2017machine,raissi2017parametric,perdikaris2017nonlinear,raissi2016deep,gulian2018machine} or a neural network \citep{RAISSI2018PINNs,JMLR:v19:18-046,raissi2018hiddenfluid,raissi2018forwardbackward,raissi2018multistep,raissi2018deepVIV}. Moreover, placing a prior on the solution itself is fully justified by the similar approach pursued in the past century by classical methods of solving partial differential equations such as finite elements, finite differences, or spectral methods, where one would expand the unknown solution in terms of an appropriate set of basis functions. However, the classical methods suffer from the curse of dimensionality mainly due to their reliance on spatio-temporal grids. In contrast, modern techniques avoid the tyranny of mesh generation, and consequently the curse of dimensionality \cite{raissi2018forwardbackward, weinan2017deep}, by approximating the unknown solution with a neural network \cite{raissi2017physics_I,raissi2017physics_II,raissi2018deep} or a Gaussian process. This transforms the problem of solving a partial differential equation into an optimization problem. This is enabling as it allows us to solve forward, backward (inverse), data-assimilation, data-driven discovery, and control problems (in addition to many other classes of problems of practical interest) using a single unified framework. On the flip side of the coin, this can help us design physics-informed learning machines.\\

\begin{figure}[t]
\centering
\includegraphics[width=0.65\textwidth]{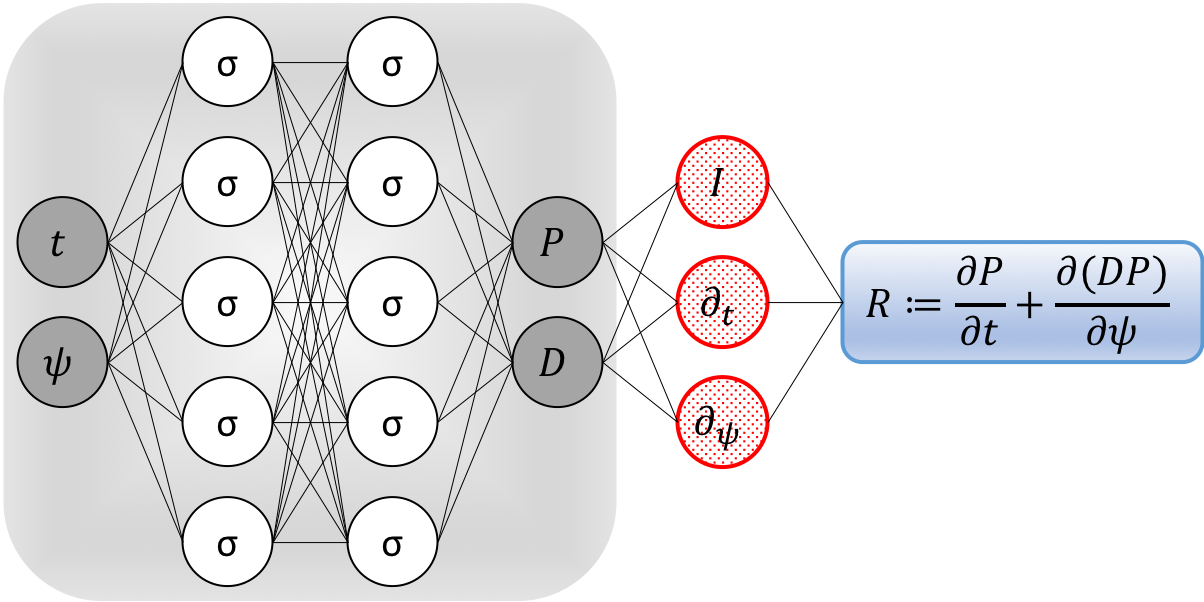}
\caption{\emph{Conditional Expected Diffusion Network:} A plain vanilla densely connected (physics uninformed) neural network, with 10 hidden layers and 50 neurons per hidden layer per output variable ({\it i.e.}, $2 \times 50 = 100$ neurons per hidden layer), takes the input variables $t$ and $\psi$ while outputting $P$ and $D$. As for the activation functions, we use $\sigma(x) = x\ \text{sigmoid}(x)$ known in the literature as \emph{Swish}. For illustration purposes only, the network depicted in this figure comprises of 2 hidden layers and 5 neurons per hidden layers. We employ automatic differentiation to obtain the required derivatives to compute the residual (physics informed) network $R$. The total loss function is composed of the regression loss of the probability density function $P$ and the loss imposed by the partial differential equation $R$. Here, $I$ denotes the identity operator and the differential operators $\partial_t$ and $\partial_\psi$ are computed using automatic differentiation and can be thought of as ``activation operators". Moreover, the gradients of the loss function are back-propagated through the entire network to train the neural network parameters using the Adam optimizer.}\label{fig:PINNs_Diffusion}
\end{figure}
The aforementioned prior assumption along with Eq.\ \eqref{eq:diffusion} will allow us to obtain the following \emph{physics informed neural network} (see Fig.\ \ref{fig:PINNs_Diffusion})
\[
R := \frac{\partial P}{\partial t} + \frac{\partial (D P)}{\partial \psi}.
\]
We obtain the required derivatives to compute the residual network $R(t,\psi)$ by applying the chain rule for differentiating compositions of functions using automatic differentiation \cite{baydin2015automatic}. It is worth emphasizing that automatic differentiation is different from, and in several respects superior to, numerical or symbolic differentiation; two commonly encountered techniques of computing derivatives. In its most basic description \cite{baydin2015automatic}, automatic differentiation relies on the fact that all numerical computations are ultimately compositions of a finite set of elementary operations for which derivatives are known. Combining the derivatives of the constituent operations through the chain rule gives the derivative of the overall composition. This allows accurate evaluation of derivatives at machine precision with ideal asymptotic efficiency and only a small constant factor of overhead. In particular, to compute the required derivatives we rely on Tensorflow \cite{abadi2016tensorflow}, which is a popular and relatively well documented open source software library for automatic differentiation and deep learning computations.\\

Parameters of the neural networks $P(t,\psi)$ and $D(t,\psi)$ can be learned by minimizing the following loss function
\[
\sum_{n=1}^N |P(t^n,\psi^n) - P^n|^2 + \sum_{n=1}^N |R(t^n,\psi^n)|^2,
\]
where $\{t^n, \psi^n, P^n\}_{n=1}^N$ represents the data on the probability density function $P(t,\psi)$. Here, the first summation corresponds to the training data on the probability density function $P(t,\psi)$ while the second summation enforces the structure imposed by Eq.\ \eqref{eq:diffusion} at a finite set of measurement points whose number and locations are taken to be the same as the training data. However, it should be pointed out that the number and locations of the points on which we enforce the set of partial differential equations could be different from the actual training data. Although not pursued in the current work, this could significantly reduce the required number of training data on the probability density function.

\subsection{Conditional Expected Dissipation}

\begin{figure}[t]
\centering
\includegraphics[width=0.7\textwidth]{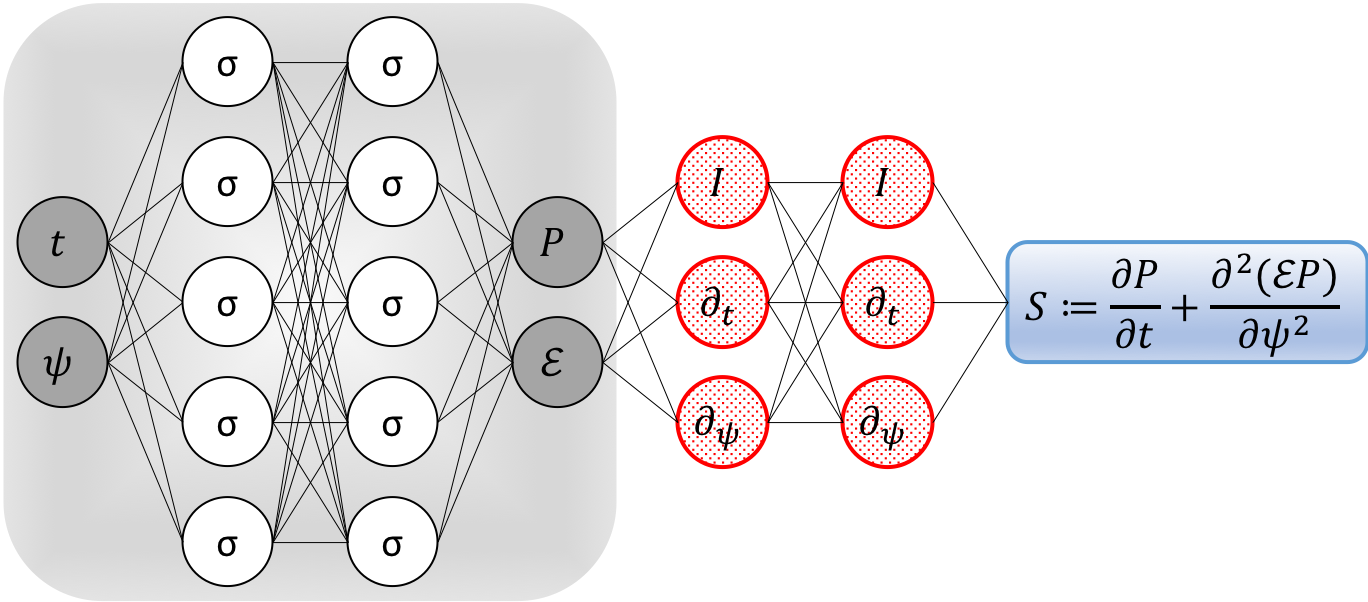}
\caption{\emph{Conditional Expected Dissipation Network:} A plain vanilla densely connected (physics uninformed) neural network, with 10 hidden layers and 50 neurons per hidden layer per output variable ({\it i.e.}, $2 \times 50 = 100$ neurons per hidden layer), takes the input variables $t$ and $\psi$ while outputting $P$ and $\calE$. As for the activation functions, we use $\sigma(x) = x\ \text{sigmoid}(x)$ known in the literature as \emph{Swish}. For illustration purposes only, the network depicted in this figure comprises of 2 hidden layers and 5 neurons per hidden layers. We employ automatic differentiation to obtain the required derivatives to compute the residual (physics informed) network $S$. If a term does not appear in the blue boxes ({\it e.g.}, $\frac{\partial^2P}{\partial t^2}$ or $\frac{\partial^2P}{\partial t\partial \psi}$), its coefficient is assumed to be zero. It is worth emphasizing that unless the coefficient in front of a term is non-zero, that term is not going to appear in the actual ``compiled" computational graph and is not going to contribute to the computational cost of a feed forward evaluation of the resulting network. The total loss function is composed of the regression loss of the probability density function $P$ and the loss imposed by the differential equation $S$. Here, $I$ denotes the identity operator and the differential operators $\partial_t$ and $\partial_\psi$ are computed using automatic differentiation and can be thought of as ``activation operators". Moreover, the gradients of the loss function are back-propagated through the entire network to train the neural network parameters using the Adam optimizer.}\label{fig:PINNs_Dissipation}
\end{figure}

Alternatively, one could proceed by approximating the function
\[
(t,\psi) \longmapsto (P, \calE)
\]
by a deep neural network taking as inputs $t$ and $\psi$ while outputting $P$ and $\calE$. This prior assumption along with Eq.\ \eqref{eq:dissipation} will allow us to obtain the following \emph{physics informed neural network} (see Fig.\  \ref{fig:PINNs_Dissipation})
\[
S := \frac{\partial P}{\partial t} + \frac{\partial^2 (\calE P)}{\partial \psi^2}.
\]
We use automatic differentiation \cite{baydin2015automatic} to acquire the required derivatives to compute the residual network $S(t,\psi)$. Parameters of the neural networks $P(t,\psi)$ and $\calE(t,\psi)$ can be learned by minimizing the following loss function
\[
\sum_{n=1}^N |P(t^n,\psi^n) - P^n|^2 + \sum_{n=1}^N |S(t^n,\psi^n)|^2,
\]
where $\{t^n, \psi^n, P^n\}_{n=1}^N$ represents the data on the probability density function $P(t,\psi)$. Here, the first summation corresponds to the training data on the probability density function $P(t,\psi)$ while the second summation enforces the structure imposed by Eq.\ \eqref{eq:dissipation} at a finite set of measurement points whose number and locations are taken to be the same as the training data.

\section{Assessment}
To assess the performance of our deep learning algorithms, we consider the amplitude mapping closure (AMC)
\cite{Kraichnan89,CCK89,Pope91}. This provides the external closure for the PDF transport in an implicit manner. This is done by {\it mapping}
of the random field of interest $\psi$ to a stationary Gaussian reference field $\psi_0$, via a transformation $\psi=\chi(\psi_0,t)$. Once this relation is established, the PDF of the random variable $\psi$, $\calP (\psi)$, is related to that of a Gaussian distribution. In a domain with fixed upper and lower bounds, the  solution for a symmetric field with zero mean, is represented in terms of an unknown time $\tau$ where
\begin{eqnarray}
    \calP (\tau, \psi) &=& \frac{\calG}{2} \exp \left\{ - (\calG^2-1) \left[{\rm erf}^{-1} (\psi)\right]^2\right\},\nonumber
   \\
    \calG (\tau) &=& \sqrt{\exp (2 \tau) -1}.\label{8}
\end{eqnarray}
The AMC captures many of the basic features of the binary mixing problem. Namely, the inverse diffusion of the PDF in the composition domain from a double delta distribution to an asymptotic approximate Gaussian distribution centered around $<\psi>$, as the variance goes to zero (or $\calG \rightarrow \infty$). There are other means of
``driving'' the PDF toward Gaussianity (or any other distribution) in a physically acceptable manner. The Johnson-Edgeworth tranlation (JET) \cite{MFMG93} involves the transformation of the random physical field $\psi$, to a fixed standard Gaussian (or any other) reference field by means of a translation of the form
\begin{eqnarray}
    &&\psi = \calZ \left[\frac{\psi_0}{\gamma (t)}\right],\nonumber\\
    &&\gamma(t=0)=0 \le \gamma (t) \le \gamma(t \rightarrow \infty)
\rightarrow \infty.\nonumber
\end{eqnarray}
The function $\gamma (t)$ here plays a role similar to that of $\calG$ in the AMC. With appropriate form for the function $\calZ$, the scalar PDF is determined.
In this manner, many frequencies can be generated. The AMC, for example, is recovered by the translation $\calZ =  {\rm erf} \left({\psi_0 / \gamma}\right)$; so can be also labeled as the ${\rm erf}^{-1}$-Normal distribution. Recognizing this translation, the relation between $\tau$ and the physical time $t$ can be determined through knowledge of the higher order statistics. For example, the normalized variance:
\begin{equation}
\frac{<\sigma^2> (\tau)}{<\sigma^2>(0)} = \frac{2}{\pi} \arctan
\left(\frac{1}{\calG \sqrt{\calG^2 + 2}} \right),
\label{11}
\end{equation}
determines $t$ through specification of the total  mean dissipation $\epsilon (t)= -\sigma \frac{d \sigma}{dt}$. With the knowledge of this dissipation, all of the conditional statistics are determined \cite{MFMG93,JGG92}
\begin{equation}
\frac{\calE (t, \psi)}{\epsilon(t)}= \left(\sqrt{ \frac{ 1 + \sin\left[ \frac{\pi
\sigma^2 (t) }{ 2 \sigma^2(0)} \right] }{ 1 - \sin\left[ \frac{\pi
\sigma^2 (t) }{ 2 \sigma^2(0)} \right]}}\right)
\exp \left\{-2 \left[{\rm erf}^{-1} ({\psi})\right]^2\right\}.
\label{33}
\end{equation}
\begin{equation}
\frac{ \calD (t, \psi) }{ \epsilon(t)} = \left( \frac{-\sqrt{\pi} }{
\sin\left[ \frac{\pi
\sigma^2 (t) }{ 2 \sigma^2(0)} \right]}
 \sqrt{ \frac{ 1 + \sin\left[ \frac{\pi
\sigma^2 (t) }{ 2 \sigma^2(0)} \right] }{ 1 - \sin\left[ \frac{\pi
\sigma^2 (t) }{ 2 \sigma^2(0)} \right]}} \right)
\exp \left\{ - \left[{\rm erf}^{-1} ({\psi})\right]^2\right\} {\rm erf}^{-1} ({\psi}).
\label{41}
\end{equation}

\section[Results]{Results\protect\footnote{All data and codes used in this manuscript will be publicly available on GitHub at https://github.com/maziarraissi/DeepTurbulence.}}\label{sec:Results}

\begin{figure}
\centering
\includegraphics[width=\textwidth]{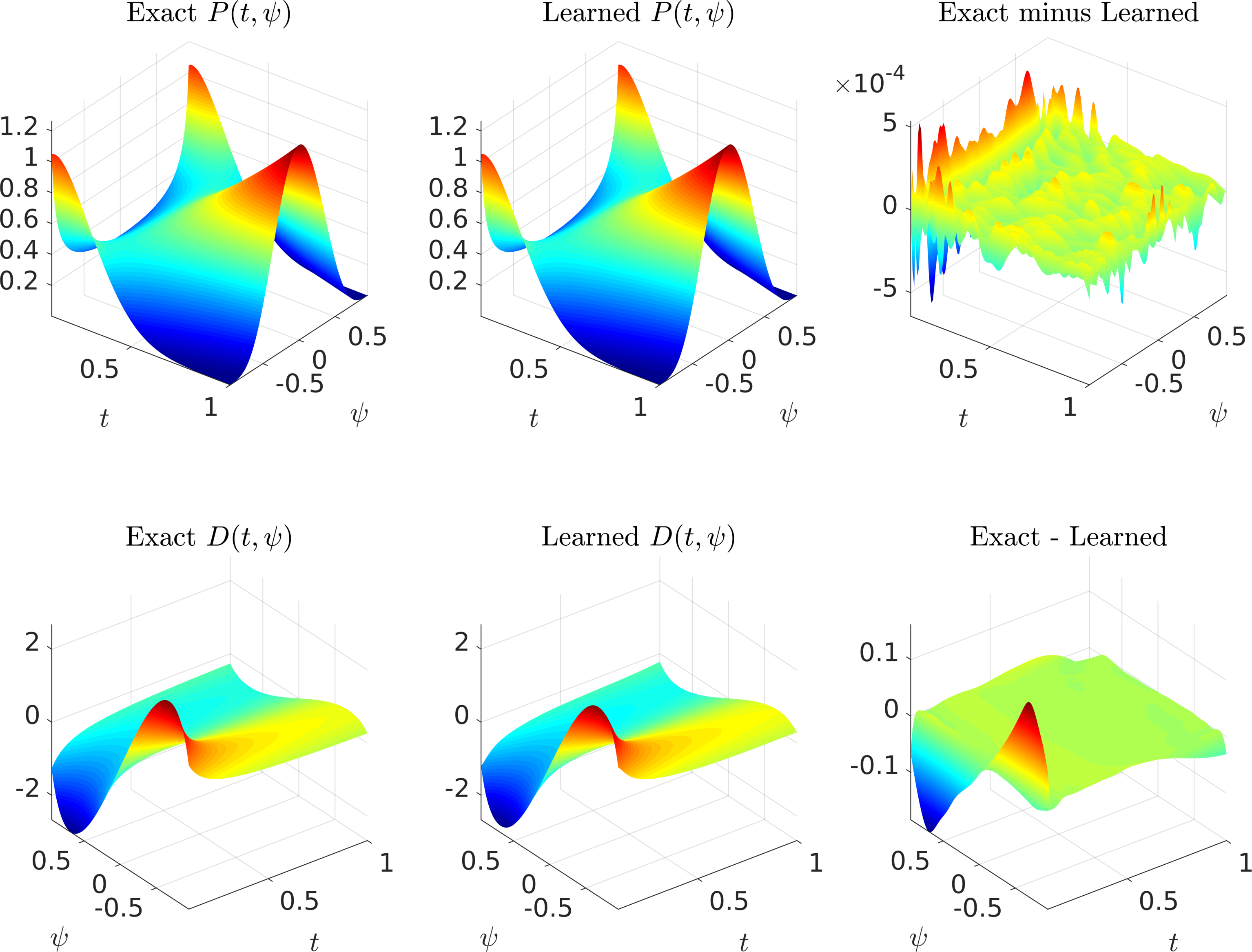}
\caption{\emph{Conditional Expected Diffusion:} The exact probability density function $P(t,\psi)$ alongside the learned one is depicted in the top panels, while the exact and learned conditional expected diffusion $D(t,\psi)$ are plotted in the bottom panels. It is worth highlighting that the algorithm has seen no data whatsoever on the diffusion coefficient.}\label{fig:turbulence_1D_diffusion}
\end{figure}

\begin{figure}
\centering
\includegraphics[width=\textwidth]{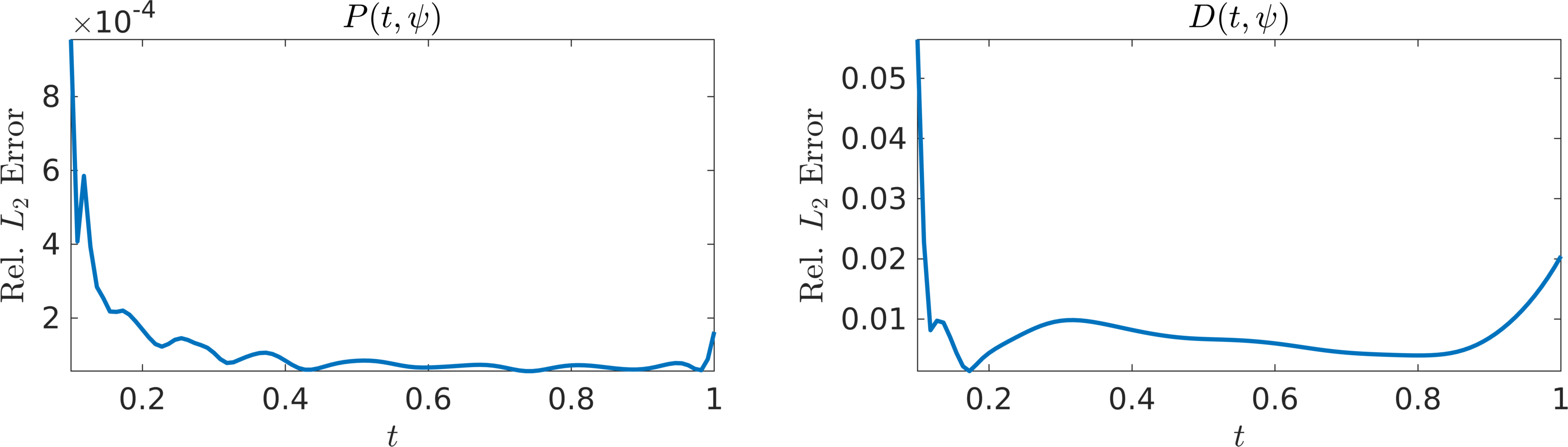}
\caption{\emph{Conditional Expected Diffusion:} The exact probability density function $P(t,\psi)$ alongside the learned one is depicted in the top panels, while the exact and learned conditional expected diffusion $D(t,\psi)$ are plotted in the bottom panels. It is worth highlighting that the algorithm has seen no data whatsoever on the diffusion coefficient.}\label{fig:turbulence_1D_diffusion_errors}
\end{figure}

In the following, the AMC (or the ${\rm erf}^{-1}$-Normal distribution) is utilized to assess the performance of our deep learning framework. In particular, Fig. \ref{fig:turbulence_1D_diffusion} depicts the exact and the learned conditional expected diffusion $D(t,\psi)$. It is worth highlighting that the algorithm has seen no data whatsoever on the conditional expected diffusion. To obtain the results reported in this figure we are approximating $P(t,\psi)$ and $D(t,\psi)$ by a deep neural network consisting of 10 hidden layers with 100 neurons per each hidden layer (see Fig.\ \ref{fig:PINNs_Diffusion}). As for the activation functions, we use $x\ \text{sigmoid}(x)$ known in the literature \cite{DBLP:journals/corr/abs-1710-05941} as the \emph{Swish} activation function. The smoothness of Swish and its similarity to ReLU make it a suitable candidate for an activation function while working with physics informed neural networks \cite{JMLR:v19:18-046}. In general, the choice of a neural network's architecture ({\it e.g.}, number of layers/neurons and form of activation functions) is crucial and in many cases still remains an art that relies on one's ability to balance the trade off between \emph{expressivity} and \emph{trainability} of the neural network \citep{raghu2016expressive}. Our empirical findings so far indicate that deeper and wider networks are usually more expressive ({\it i.e.}, they can capture a larger class of functions) but are often more costly to train ({\it i.e.}, a feed-forward evaluation of the neural network takes more time and the optimizer requires more iterations to converge). In this work, we have tried to choose the neural networks' architectures in a consistent fashion throughout the manuscript by setting the number of hidden layers to 10 and the number of neurons to 50 per output variable. Consequently, there might exist other architectures that improve some of the results reported here.\\

As for the training procedure, our experience so far indicates that while training deep neural networks, it is often useful to reduce the learning rate as the training progresses. Specifically, the results reported here are obtained after $10^5$, $2 \times 10^5$, $3 \times 10^5$, and $4 \times 10^5$ consecutive epochs of the Adam optimizer \cite{kingma2014adam} with learning rates of $10^{-3}$, $10^{-4}$, $10^{-5}$, and $10^{-6}$, respectively. Each epoch corresponds to one pass through the entire dataset. The total number of iterations of the Adam optimizer is therefore given by $10^6$ times the number of data divided by the mini-batch size. The mini-batch size we used is $20000$ and the number of data points is $20000$. Every $10$ iterations of the optimizer takes around $0.04$ on a single NVIDIA Titan X GPU card. The algorithm is capable of reconstructing the probability density function $P(t,\psi)$ as well as the unknown conditional expected diffusion $D(t,\psi)$ with relative $L_2$ errors of $1.27\times 10^{-4}$ and $1.73 \times 10^{-2}$, respectively. The relative $L_2$ errors in space as a function of time are depicted in Fig.\ \ref{fig:turbulence_1D_diffusion_errors}. The relative error is high at small $t$ and that is due to the singularity of $P(t,\psi)$ at $t=0$. However, at larger times, the error decreases as the effect of initial singularity weakens.\\


\begin{figure}
\centering
\includegraphics[width=\textwidth]{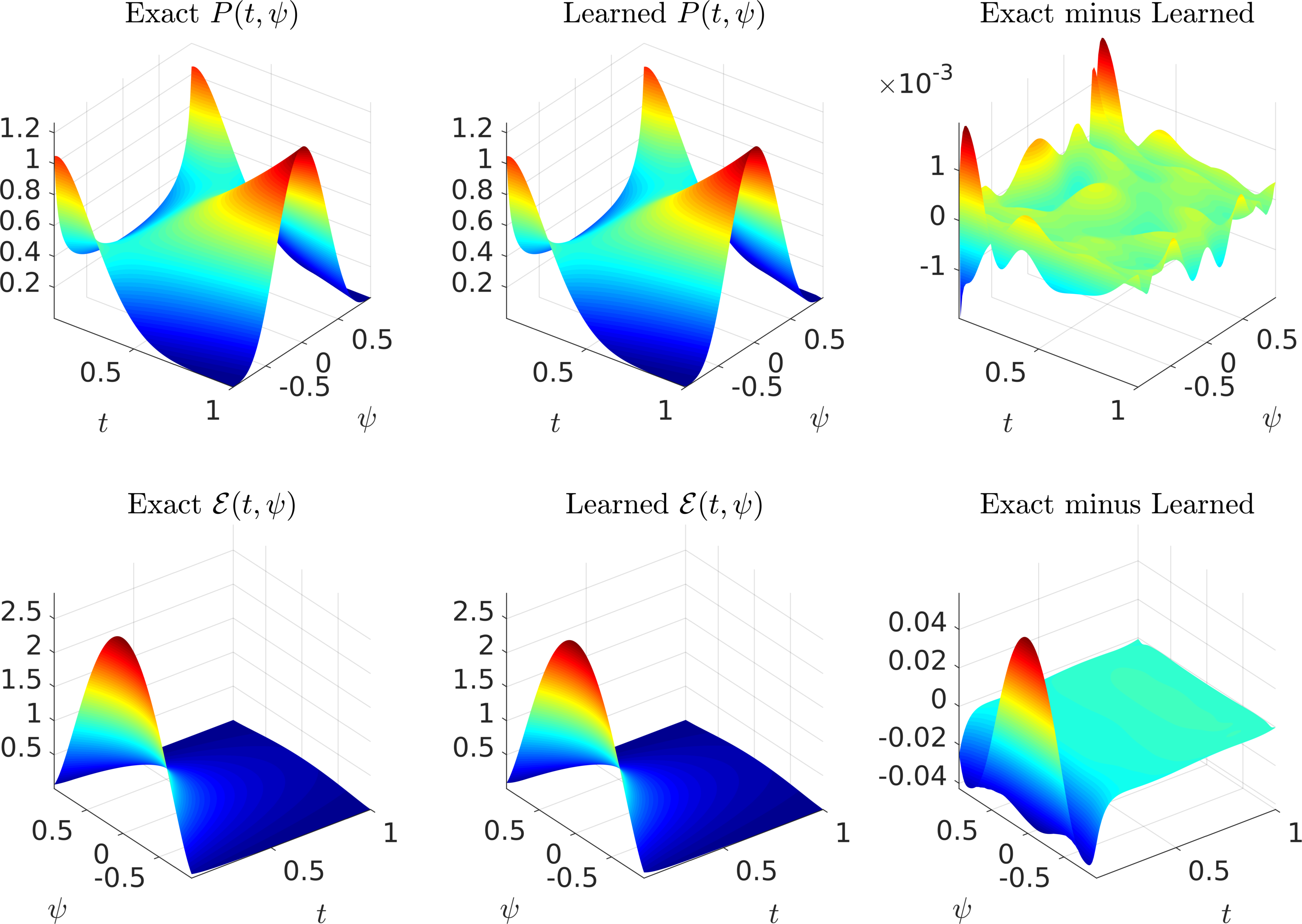}
\caption{\emph{Conditional Expected Dissipation:} The exact probability density function $P(t,\psi)$ alongside the learned one is depicted in the top panels, while the exact and learned conditional expected dissipation $\calE(t,\psi)$ are plotted in the bottom panels. It is worth highlighting that the algorithm has seen no data whatsoever on the dissipation coefficient.}\label{fig:turbulence_1D_dissipation}
\end{figure}

\begin{figure}
\centering
\includegraphics[width=\textwidth]{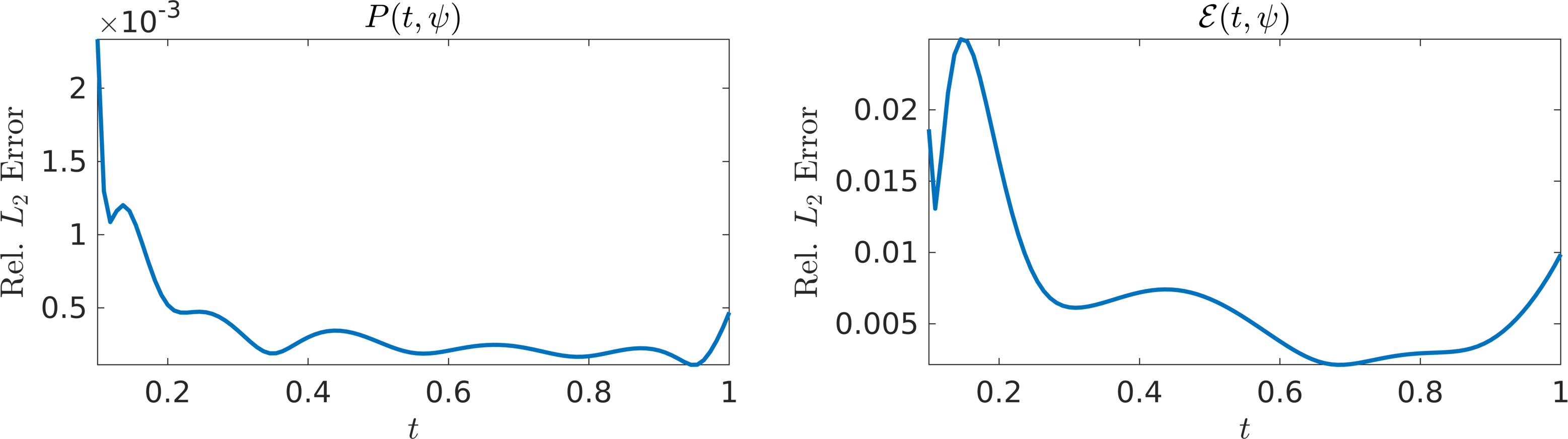}
\caption{\emph{Conditional Expected Dissipation:} The exact probability density function $P(t,\psi)$ alongside the learned one is depicted in the top panels, while the exact and learned conditional expected dissipation $\calE(t,\psi)$ are plotted in the bottom panels. It is worth highlighting that the algorithm has seen no data whatsoever on the dissipation coefficient.}\label{fig:turbulence_1D_dissipation_errors}
\end{figure}

Figure \ref{fig:turbulence_1D_dissipation} depicts the exact and the learned conditional expected dissipation $\calE(t,\psi)$. It is worth highlighting that the algorithm has seen no data whatsoever on the dissipation coefficient. To obtain the results reported in this figure we are approximating $P(t,\psi)$ and $\calE(t,\psi)$ by a deep neural network outputting two variables consisting of 10 hidden layers with 100 neurons per each hidden layer  (see Fig. \ref{fig:PINNs_Dissipation}). As for the activation functions, we use $x\ \text{sigmoid}(x)$. The training procedure is the same as the one explained above, while every $10$ iterations of the optimizer takes around $0.07$. The algorithm is capable of reconstructing the probability density function $P(t,\psi)$ as well as the unknown conditional expected dissipation $\calE(t,\psi)$ with relative $L_2$ errors of $3.84\times 10^{-4}$ and $1.75 \times 10^{-2}$, respectively. The relative $L_2$ errors in space as a function of time are depicted in Fig.\ \ref{fig:turbulence_1D_dissipation_errors}.

\section{Concluding Remarks}
In this paper we present a data-driven framework for learning unclosed terms for turbulent scalar mixing. In the presented framework the unclosed terms  are learned by (i) incorporating the physics, {\it i.e.} the PDF transport equation, and (ii) observe some high-fidelity observations on the PDF.  
We envision that the presented framework as described above can be straightforwardly extended to high-dimensional cases involving the mixing of multiple species. Early evidence of this claim can be found in \cite{raissi2018forwardbackward, weinan2017deep}, in which the authors circumvents the tyranny of numerical discretization and devise algorithms that are scalable to high-dimensions. A similar technique can be applied here while taking advantage of the fact that the data points $\{t^n, \psi^n, P^n\}_{n=1}^N$ lie on a low dimensional manifold simply because $\psi(t,x)$ is a function from a low dimensional space ({\it i.e.}, $(t,x)$) to the possibly high-dimensional space of species $\psi$. Moreover, the approach advocated in the current work is also highly scalable to the big data regimes routinely encountered while studying turbulence simply because the data will be processed in mini-batches.

\section*{Acknowledgements}
The work at Brown University is supported by the DARPA EQUiPS grant N66001-15-2-4055  and by the AFOSR Grant FA9550-17-1-0013. All data and codes used in this manuscript will be publicly available on GitHub at \url{https://github.com/maziarraiss/DeepTurbulence}.





\bibliographystyle{model1-num-names}
\bibliography{sample.bib,cfd_peyman,cfd,Hessam}







\end{document}